\documentclass[11pt]{article}
\usepackage{hyperref}
\pdfoutput=1
\begin{document}
\title{Knotted Vortices: Entropic Lattice Boltzmann Method for Vortex Dynamics}
\author{ \large $^1$Chikatamarla S. S., $^2$Favre J., $^1$Boesch F., $^1$Karlin I.V., \\
\\\vspace{6pt}\small $^1$LAV, Institute of Energy Technology,   
              \small ETH Zurich, Switzerland. \\
				      \small $^2$Swiss National Supercomputing Center CSCS, 
              \small Lugano, Switzerland. }
\maketitle
 
\begin{abstract}
This fluid dynamics video shows "knotted" vortices in real fluids. 
\end{abstract}

Although the existence of Knots and links in a real fluid was conjectured and speculated for over a century, no evidence existed until the recent experimental work of Kelckner and Irvine \cite{KI1}. Following closely the experimental setup of \cite{KI1} we present here a simulation of two linked ring vortices tied within each other. The vortex ring pair goes on to produce a reconnection event that is well captured in the simulation. The simulation was run at Re$=40'000$ based on the airfoil chord (the cross-section of the ring) and the velocity of the rings. The chord of the airfoil was resolved with $80$ grid points and the global simulation domain was around $1000^3$ grid points.

Numerous advantages of the simulation method used, the entropic lattice Boltzmann method (ELBM), such as ease of handling complex geometries\cite{CKBC}, easy parallelism and built in sub-grid features\cite{KFO,CKPRL} are evident in the video. The run time for this setup was around 8 hours using 2048 CPUs (cores) of the Monte Rosa machine located at CSCS, Lugano, Switzerland. Enhanced stability and lack of tuning parameters makes ELBM ideally suited for investigating turbulent flow phenomenon\cite{JFM}.

Click for 
\href{anc/ELBM_Vortex_rings_Hi_res.mp4}{hi resolution Video}
 and 
\href{anc/ELBM_Vortex_rings_Lo_res.mpg}{low resolution Video}.

\end{document}